\documentclass[final]{aipproc}
\layoutstyle{6x9}
\usepackage{gensymb}

\input babarsym
\def\etal{{\em et al.}}

\def\dbline{\noalign{\vskip 0.15truecm\hrule}\noalign{\vskip 2pt}\noalign{\hrule\vskip 0.15truecm}}

\newcommand{\DE}{\ensuremath{\Delta E}}
\newcommand{\dt}{\ensuremath{\Delta t}}

\def\babar{{\em B}{\footnotesize\em A}{\em B}{\footnotesize\em AR}}

\def\gevcc{\mbox{${\mathrm{GeV}}/c^2\ $}}

\def\Bz{\mbox{$\overline {B^{0}}\ $}}

\def\BB{\mbox{$\B^{+}\  \B^{-}\ $}}

\def\qqbar{\mbox{$q\bar q\ $}}

\def\Bz{\mbox{$B^0\ $}}
\def\Bzb{\mbox{$\overline{B}^0\ $}}
\def\beq{\begin{equation}}
\def\eeq{\end{equation}}
\def\bef{\begin{figure}}
\def\edf{\end{figure}}
\def\ben{\begin{enumerate}}
\def\een{\end{enumerate}}
\def\bear{\begin{array}}
\def\enar{\end{array}}
\def\beqa{\begin{eqnarray}}
\def\eeqa{\end{eqnarray}}
\def\to{\mbox{$\rightarrow$}}

\def\gevcc{\mbox{${\mathrm{GeV}}/c^2$}}

\def\FourS{\mbox{$\Upsilon{\mathrm( 4S)}$}}

\def\Bz{\mbox{${B^{0}}$}}
\def\Bzb{\mbox{$\overline B^{0}$}}
\def\BB{\mbox{$B\overline B$}}
\def\BpBm{\mbox{$B^{+} B^{-}$}}

\def\piz{\mbox{${\pi^{0}}$}}
\def\pip{\mbox{${\pi^{+}}$}}
\def\pim{\mbox{${\pi^{-}}$}}

\def\pimp{\mbox{${\pi^{\mp}}$}}

\def\Kp{\mbox{${K^{+}}$}}
\def\Km{\mbox{${K^{-}}$}}

\def\DE{\mbox{${\Delta E}$}}

\def\pipi     {\ensuremath{\pi\pi}}

\def\tcp {\ensuremath{t_{\CP}}}
\def\ttag {\ensuremath{t_{\rm tag}}}

\newcommand{\Sf}{\ensuremath{S_f}}
\newcommand{\Cf}{\ensuremath{C_f}}

\begin{document}

\title{Measurements of Time-Dependent \CP\ Asymmetries in $\boldmath{b\to s}$
  Penguin Dominated Hadronic $B$ Decays at \babar}

\classification{13.25.Hw, 12.15.Hh, 11.30.Er, 13.66.Bc,
                14.40.Cs, 13.25.Gv, 13.25.Jx, 13.20.Jf.}
\keywords      {Charmless Hadronic $B$ decays, Time-Dependent \CP\
                Violation, \stwob\ Measurement.}

\author{Pietro Biassoni\\ (On behalf of the \babar\ Collaboration)}{
  address={Universit\`a degli Studi and INFN Milano, via Celoria 16, 
    I-20133 Milano, Italy}
}

\begin{abstract}
We report measurements of Time-Dependent \CP\ asymmetries in several
$b\to s$ penguin dominated hadronic $B$ decays, where New Physics
contributions may appear.  We find no significant discrepancies with
respect to the Standard Model expectations.
\end{abstract}

\maketitle

\section{Introduction}

The measurement of \CP\ violation in $B$ meson decays  provides crucial tests of the Standard
Model (SM) and of the Cabibbo-Kobayashi-Maskawa (CKM) mechanism~\cite{CKM}.

%b->s decays
CKM-suppressed  $b\to \qqbar s$ ($q=u,d,s$) processes are dominated by
a single loop (penguin) amplitude, that, assuming penguin dominance
and neglecting higher order contributions, is expected to have the
same phase $\beta$ of the CKM-favored $b\rightarrow c \bar{c} s$
transition~\cite{ccs}.  
In many extensions of the SM  new heavy particles  may appear in the loop
\cite{Penguin}, giving rise to deviations from this expectation.
These deviations are expected to be channel dependent. 
The measurement of the phase difference between
$\Bz\to K^{*}(892)^+\pim$ and $\Bzb\to K^*(892)^-\pip$ can be used to
constrain the CKM parameters in the $(\bar{\rho},\bar{\eta})$ 
plane~\cite{gamma}.

\section{Time-Dependent Decay Rates}
The CKM phase $\beta$ is accessible experimentally through
the interference between the decay of mixed and unmixed $B$ meson into a 
\CP\ eigenstate.  This interference is observable through the time
evolution of the decay. 

In the studies reported in this  presentation, one \Bz\ from
$\FourS\ra\BzBzb$  is reconstructed in  
$\eta^{\prime}\KS$, $\eta^{\prime}\KL$, $\omega\KS$, or $\KS\!\KS\!\KS$ \CP\
eigenstate, or in \pip\pim\KS\ or \Kp\Km\KS\ non-\CP\ eigenstate final
state ($B_{sig}$), and its vertex fitted using all charged daughter tracks. 
In \KS\KS\KS\ mode, where no charged track is present at $B^0$ meson
decay vertex,  $B_{sig}$ vertex is identified using the $\KS$
reconstructed flight directions and the knowledge of the average
interaction point~\cite{IP}.  
From the remaining particles in the
event we reconstruct the decay vertex of the other $B$ meson ($B_{\rm
  tag}$) and identify its flavor, through the analysis of the
decay product of $B_{\rm tag}$~\cite{Tagging}. 
  
The distribution of the difference
$\deltat \equiv \tcp - \ttag$ of the proper decay times 
of $B$ mesons into \CP-eigenstate final states is given by
\begin{eqnarray}
 f(\dt) &=& 
 \frac{e^{-\left|\deltat\right|/\tau}}{4\tau} \{1 \pm 
\left[-\eta_f \Sf\sin(\deltamd\deltat) -
  \Cf\cos(\deltamd\deltat)\right]\}
\label{eq:td}
\end{eqnarray}
where $\eta_f$ is the \CP\ eigenvalue of the final state $f$ and
$\tau$ is the \Bz\  meson  lifetime.  The upper
(lower) sign denotes a decay accompanied by a \Bz (\Bzb) tag, and
$\deltamd$ is the mixing frequency.  

For three body  non-\CP-eigenstate final state, the
\CP-violating parameters are a function of  
the position over the Dalitz Plot (DP). In this case
Eq. ~(\ref{eq:td}) is written as
\begin{eqnarray}
 f(\dt) &=& 
 \frac{e^{-\left|\deltat\right|/\tau}}{4\tau} \left\{|A|^2 +|\overline{A}|^2 \pm 
   \left[\eta_f2Im[\overline{A}A^* ]\sin(\deltamd\deltat)
     -\right.\right.\nonumber\\
&&\left.\left.
 (|A|^2-|\overline{A}|^2)\cos(\deltamd\deltat)
\right]\right\}.
\label{eq:DPtp}
\end{eqnarray}
Let the decay $B^0\to X_1X_2X_3$  proceed through $N$ intermediate 
states: 
the amplitude $A$ depends only on the Mandelstam invariants $s_{12}$
and $s_{23}$, 
and in the isobar approximation is
\begin{eqnarray}
A(s_{12},s_{23}) = \sum_{j=1}^N
|c_j|e^{-i\phi_j}R_j(m_j)X_L(|\vec{p}*|r^{\prime})
X_L(|\vec{q}|r)T_j(L,\vec{p},\vec{q})
\end{eqnarray}
where $c_j$ and $\phi_j$ are the relative magnitude and phase of 
the decay mode $j$, $R_j(m)$ is the lineshape term, $X_L$ are 
Blatt-Weisskopf barrier factors~\cite{barrier}, $T_j$ is the angular
distribution, $\vec{p}$ ($\vec{q}$) is the momentum of the prompt
particle (one of the resonance daughters), $L$ is the orbital
angular momentum between $\vec{p}$ and the resonance momentum, and
asterisk denotes $B$ rest frame. 
For a decay into a quasi-two-body \CP\ eigenstate, one can extract the
parameters 
$\beta_{eff} = \frac{1}{2}\arg(c_k\bar{c}_k^*)$ and
$\calA_{ch}(k)=[|\bar{c}_k|^2-|c_k|^2]/[|\bar{c}_k|^2+|c_k|^2]$.  
For a decay into quasi-two-body non-\CP\ eigenstate, we measure the charge
asymmetry and the phase between the two conjugate
states $\Delta\Phi(k)=\arg(c_k\bar{c}_k^*)$.

%deviation of S and C from 0
A nonzero value of the parameter \Cf\ or $\calA_{ch}$ would indicate
direct \CP\ violation. 
In these modes we expect 
$-\eta_f\Sf \equiv  -\eta_f\sin2\beta_{eff}\approx\stwob$.
Deviations $\Delta S_f=S_f - \stwob$ from this expectation may appear
even within the SM~\cite{Gross,london}, and are estimated in
several theoretical approaches~\cite{Gross,beneke}. 

\section{Analysis Technique}

Analyses presented here are based on a sample of
$465\times10^6$ \BB\ pairs ($383\times10^6$ for $\Bz\to\KS\pip\pim$),
collected at a center-of-mass energy equal to the mass of the \FourS\
resonance at the PEP-II asymmetric \epem\ collider, at the SLAC
National Accelerator Laboratory, and recorded by the \babar\
detector~\cite{BABARNIM}. 
The $B$ meson is reconstructed into the above-mentioned \CP\ eigenstates.
The $B$ meson is kinematically characterized by  the variables
$\Delta E\equiv E_B-\frac{1}{2}\sqrt{s}$ and $\mes \equiv \sqrt{s/4 -
|\vec{p}_B|^2}$, where $(E_B,\vec{p}_B)$ is the $B$ four-momentum vector
expressed in \FourS\ rest frame.

Background arises primarily from random combinations of particles in
$\epem\to\qqbar$ events ($q=u,d,s,c$).
We suppress this background with requirements on the event shape variables
and on the energy, invariant mass and particle identification
signature of the decay products.
All events are required to have $|\dt|<20$~ps and $\sigma_{\Delta
  t}<2.5$~ps. 

%\CP-violation, DP parameters and signal yields 
For each mode, results are obtained from an extended maximum likelihood
fit with input variables $\Delta 
E$, \mes, \deltat, and the output of a multivariate discriminant
combining different event shape variables. 
In $\omega\KS$ decay we also use $\omega$ mass and angular variables into
the fit.
$\KL$ momentum is determined using a $B$ mass constraint, hence \mes\
is fully correlated to \DE, and is not used into the fit in
$\eta^{\prime}\KL$ modes.  
The likelihood for a given event is the sum of the signal, continuum
and the $B$-background components, weighted by their respective event
yields. 
In $\KS\pip\pim$ and $\KS K^+K^-$ modes, a time-dependent DP analysis
is performed. The DP model includes $f_0(980)$, $\rho^0(770)$,
$K^{*\pm}(892)$, $(K\pi)_0^{*\pm}$, $f_2(1240)$, $f_x(1300)$,
$\chi_{c0}$ ($f_0(980)$, $\phi(1020)$, $X(1550)$, $f_2(1270)$,
$\chi_{c0}$, $D^\pm$, $D_s^\pm$) and non resonant component for
$\KS\pip\pim$ ($\KS K^+ K^-$) decay mode.
In $\KS K^+ K^-$ analysis, the fit is first performed on the whole DP,
and then in the low (high) mass region $m_{K^+K^-}<1.1$~\gevcc\
($m_{K^+K^-}>1.1$~\gevcc), fixing all the parameters to the values
found in the whole DP fit, except the ones involving the $f_0(980)$
($\phi(1020)$) resonance.

\section{Results}
In Table~\ref{tab:res} and~\ref{tab:resDP} we report the results for
\CP-violating parameters in analyses of the decay of a  \Bz\ meson
into a \CP\ eigenstates and a three body non-\CP\ eigenstates final
state (DP analyses), respectively~\cite{etaPK}. 
Results for $\KS K^+ K^-$ and \KS\KS\KS are preliminary.\\

\begin{table}[htb]
\caption{Results of analyses of $b\to s$ decays into \CP\
  eigenstates. For each decay mode we report $-\eta_fS_f$ and
  $C_f$. The first error is 
  statistical, the second systematic.}
\begin{tabular}{l|c|c|c}
\dbline
Decay Mode & $-\eta_fS_f$ & $C_f$ \\
\hline
$\eta^{\prime}K^0$ &  $0.57\pm0.08\pm0.02$&
$-0.08\pm0.06\pm0.02$\\
\hline
$\omega\KS$ & $0.55^{+0.26}_{-0.29}\pm0.02$ &
$-0.52^{+0.22}_{-0.20}\pm0.03$\\
\hline
$\KS\KS\KS$  &$0.90^{+0.20\;+0.04}_{-0.18\;-0.03}$&
$-0.16\pm0.17\pm0.03$\\ 
\dbline
\end{tabular}
\label{tab:res}
\end{table}

In \KS\pip\pim and $\KS K^+ K^-$ low mass region, the likelihood
function has two minima. In $\Bz\to f_0(980)\KS$ with $f_0(980)\to
K^+K^-$, the second solution 
is disfavored by the result from
$f_0(980)\to\pip\pim$.
In \KS\pip\pim\ analysis we measure
$\calA_{ch}(K^*(892)^+\pim)=0.20\pm0.10\pm0.02$, where the first (second) error
is statistical (systematic). We also exclude
$-137\degree<\Delta\Phi(K^*(892)^+\pim)<-5\degree$ at 95\% confidence level. 

\section{Conclusions}
We have reported the results of measurements of \CP-violating
parameters in several $b\to s$ hadronic $B$ meson decays. All the
results are consistent with the SM. Results are in agreement with and
supersede previous \babar\ measurements.\\

\begin{table}[!h]
\caption{Results of DP $b\to s$ analyses. For each decay mode we report
  $\beta_{eff}$, and $\calA_{ch}$, for both
  solutions. The first error is
  statistical, the second systematic.}
\begin{tabular}{l|c|c|c|c}
\dbline
Decay Mode & \multicolumn{2}{|c|}{Solution I} &
\multicolumn{2}{|c}{Solution II}\\ 
 & $\beta_{eff}$ ($\degree$) & $\calA_{ch}$ &$\beta_{eff}$
($\degree$)& $\calA_{ch}$\\ 
\hline
\multicolumn{5}{c}{\KS\pip\pim}\\
\hline
$f_0(980)\KS$& $36.0\pm9.8\pm3.0$ &$-0.08\pm0.19\pm0.05$
&$56.2\pm10.4\pm3.0$&$-0.23\pm0.19\pm0.05$\\ 
$\rho^0(770)\KS$& $10.2\pm8.9\pm3.6$&$0.05\pm0.26\pm0.10$&$33.4\pm10.4\pm3.6$&$0.14\pm0.26\pm0.10$\\
\hline
\multicolumn{5}{c}{$\KS K^+K^-$}\\
\hline
Whole DP & $25.2\pm4.0\pm1.1$ & $0.03\pm0.07\pm0.02$ & -- & -- \\
High Mass & $29.8\pm4.6\pm1.7$ & $0.05\pm0.09\pm0.04$ & -- & -- \\
$\phi$\KS & $7.4\pm7.4\pm1.1$ &$0.14\pm0.19\pm0.02$ &
$8.0\pm8.0\pm1.1$ & $0.13\pm0.18\pm0.02$\\ 
$f_0(980)\KS$& $8.6\pm7.4\pm1.7$ & $0.01\pm0.26\pm0.07$
&$197.1\pm10.9\pm1.7$& $-0.49\pm0.25\pm0.07$ \\
\dbline
\end{tabular}
\label{tab:resDP}
\end{table}

\section{Acknowledgments}
I'd like to thank all my \babar\ colleagues for their support and in
particular Fernando Palombo and Alfio Lazzaro.

\bibliographystyle{aipprocl} % if natbib is missing

\end{document}